\documentclass[twocolumn,tight]{aastex631}
\graphicspath{{./}{figures/}}
\usepackage[T1]{fontenc}

\begin{document}

\title{exoALMA I. Science Goals, Project Design and Data Products}

\author[0000-0003-1534-5186]{Richard Teague}
\affiliation{Department of Earth, Atmospheric, and Planetary Sciences, Massachusetts Institute of Technology, Cambridge, MA 02139, USA}

\author[0000-0000-0000-0000]{Myriam Benisty}
\affiliation{Universit\'e C\^ote d'Azur, Observatoire de la C\^ote d'Azur, CNRS, Laboratoire Lagrange, France}
\affiliation{Max-Planck Institute for Astronomy (MPIA), Königstuhl 17, 69117 Heidelberg, Germany}

\author[0000-0003-4689-2684]{Stefano Facchini}
\affiliation{Dipartimento di Fisica, Universit\`a degli Studi di Milano, Via Celoria 16, 20133 Milano, Italy}

\author[0000-0003-1117-9213]{Misato Fukagawa} 
\affiliation{National Astronomical Observatory of Japan, Osawa 2-21-1, Mitaka, Tokyo 181-8588, Japan}

\author[0000-0001-5907-5179]{Christophe Pinte}
\affiliation{Univ. Grenoble Alpes, CNRS, IPAG, B38000 Grenoble, France}
\affiliation{School of Physics and Astronomy, Monash University, Clayton VIC 3800, Australia}

\author[0000-0003-2253-2270]{Sean M. Andrews}
\affiliation{Center for Astrophysics | Harvard \& Smithsonian, Cambridge, MA 02138, USA}

\author[0000-0001-7258-770X]{Jaehan Bae}
\affiliation{Department of Astronomy, University of Florida, Gainesville, FL 32611, USA}

\author[0000-0001-6378-7873]{Marcelo Barraza-Alfaro}
\affiliation{Department of Earth, Atmospheric, and Planetary Sciences, Massachusetts Institute of Technology, Cambridge, MA 02139, USA}

\author[0000-0002-2700-9676]{Gianni Cataldi} 
\affiliation{National Astronomical Observatory of Japan, 2-21-1 Osawa, Mitaka, Tokyo 181-8588, Japan}

\author[0000-0003-3713-8073]{Nicolás Cuello} 
\affiliation{Univ. Grenoble Alpes, CNRS, IPAG, 38000 Grenoble, France}

\author[0000-0003-2045-2154]{Pietro Curone} 
\affiliation{Dipartimento di Fisica, Universit\`a degli Studi di Milano, Via Celoria 16, 20133 Milano, Italy}
\affiliation{Departamento de Astronom\'ia, Universidad de Chile, Camino El Observatorio 1515, Las Condes, Santiago, Chile}

\author[0000-0002-1483-8811]{Ian Czekala} 
\affiliation{School of Physics \& Astronomy, University of St. Andrews, North Haugh, St. Andrews KY16 9SS, UK}
\affiliation{Centre for Exoplanet Science, University of St. Andrews, North Haugh, St. Andrews, KY16 9SS, UK}

\author[0000-0003-4679-4072]{Daniele Fasano} 
\affiliation{Laboratoire Lagrange, Université Côte d’Azur, CNRS, Observatoire de la Côte d’Azur, 06304 Nice, France}
\affiliation{Univ. Grenoble Alpes, CNRS, IPAG, 38000 Grenoble, France}

\author[0000-0002-9298-3029]{Mario Flock} 
\affiliation{Max-Planck Institute for Astronomy (MPIA), Königstuhl 17, 69117 Heidelberg, Germany}

\author[0000-0002-5503-5476]{Maria Galloway-Sprietsma}
\affiliation{Department of Astronomy, University of Florida, Gainesville, FL 32611, USA}

\author[0009-0003-8984-2094]{Charles H. Gardner}
\affiliation{Department of Physics and Astronomy, Rice University,
Houston, TX 77005, USA}
\affiliation{Los Alamos National Laboratory, Los Alamos, NM 87545, USA}

\author[0000-0002-5910-4598]{Himanshi Garg}
\affiliation{School of Physics and Astronomy, Monash University, Clayton VIC 3800, Australia}

\author[0000-0002-8138-0425]{Cassandra Hall} 
\affiliation{Department of Physics and Astronomy, The University of Georgia, Athens, GA 30602, USA}
\affiliation{Center for Simulational Physics, The University of Georgia, Athens, GA 30602, USA}
\affiliation{Institute for Artificial Intelligence, The University of Georgia, Athens, GA, 30602, USA}

\author[0000-0003-1502-4315]{Iain Hammond} 
\affiliation{School of Physics and Astronomy, Monash University, VIC 3800, Australia}

\author[0000-0001-7641-5235]{Thomas Hilder} 
\affiliation{School of Physics and Astronomy, Monash University, VIC 3800, Australia}

\author[0000-0001-6947-6072]{Jane Huang} 
\affiliation{Department of Astronomy, Columbia University, 538 W. 120th Street, Pupin Hall, New York, NY, USA}

\author[0000-0003-1008-1142]{John~D.~Ilee} 
\affiliation{School of Physics and Astronomy, University of Leeds, Leeds, UK, LS2 9JT}

\author[0000-0001-8061-2207]{Andrea Isella}
\affiliation{Department of Physics and Astronomy, Rice University, 6100 Main St, Houston, TX 77005, USA}
\affiliation{Rice Space Institute, Rice University, 6100 Main St, Houston, TX 77005, USA}

\author[0000-0001-8446-3026]{Andr\'es F. Izquierdo} 
\affiliation{Department of Astronomy, University of Florida, Gainesville, FL 32611, USA}
\affiliation{Leiden Observatory, Leiden University, P.O. Box 9513, NL-2300 RA Leiden, The Netherlands}
\affiliation{European Southern Observatory, Karl-Schwarzschild-Str. 2, D-85748 Garching bei M\"unchen, Germany}
\affiliation{NASA Hubble Fellowship Program Sagan Fellow}

\author[0000-0001-7235-2417]{Kazuhiro Kanagawa} 
\affiliation{College of Science, Ibaraki University, 2-1-1 Bunkyo, Mito, Ibaraki 310-8512, Japan}

\author[0000-0002-8896-9435]{Geoffroy Lesur} 
\affiliation{Univ. Grenoble Alpes, CNRS, IPAG, 38000 Grenoble, France}

\author[0000-0002-2357-7692]{Giuseppe Lodato} 
\affiliation{Dipartimento di Fisica, Universit\`a degli Studi di Milano, Via Celoria 16, 20133 Milano, Italy}

\author[0000-0003-4663-0318]{Cristiano Longarini} 
\affiliation{Institute of Astronomy, University of Cambridge, Madingley Road, CB3 0HA, Cambridge, UK}
\affiliation{Dipartimento di Fisica, Universit\`a degli Studi di Milano, Via Celoria 16, 20133 Milano, Italy}

\author[0000-0002-8932-1219]{Ryan A. Loomis}
\affiliation{National Radio Astronomy Observatory, 520 Edgemont Rd., Charlottesville, VA 22903, USA}

\author[0000-0002-9626-2210]{Fr\'ed\'eric Masset}
\affiliation{Instituto de Ciencias F\'isicas, Universidad Nacional Aut\'onoma de M\'exico, Av. Universidad s/n, 62210 Cuernavaca, Mor., Mexico}

\author[0000-0002-1637-7393]{Francois Menard} 
\affiliation{Univ. Grenoble Alpes, CNRS, IPAG, 38000 Grenoble, France}

\author[0000-0003-4039-8933]{Ryuta Orihara} 
\affiliation{College of Science, Ibaraki University, 2-1-1 Bunkyo, Mito, Ibaraki 310-8512, Japan}

\author[0000-0002-4716-4235]{Daniel J. Price} 
\affiliation{School of Physics and Astronomy, Monash University, Clayton VIC 3800, Australia}

\author[0000-0003-4853-5736]{Giovanni Rosotti} 
\affiliation{Dipartimento di Fisica, Universit\`a degli Studi di Milano, Via Celoria 16, 20133 Milano, Italy}

\author[0000-0002-0491-143X]{Jochen Stadler} 
\affiliation{Universit\'e C\^ote d'Azur, Observatoire de la C\^ote d'Azur, CNRS, Laboratoire Lagrange, 06304 Nice, France}
\affiliation{Univ. Grenoble Alpes, CNRS, IPAG, 38000 Grenoble, France}

\author[0000-0003-1859-3070]{Leonardo Testi}
\affiliation{Alma Mater Studiorum Università di Bologna, Dipartimento di Fisica e Astronomia (DIFA), Via Gobetti 93/2, 40129 Bologna, Italy}
\affiliation{INAF-Osservatorio Astrofisico di Arcetri, Largo E. Fermi 5, 50125 Firenze, Italy}

\author[0000-0003-1412-893X]{Hsi-Wei Yen} 
\affiliation{Academia Sinica Institute of Astronomy \& Astrophysics, 11F of Astronomy-Mathematics Building, AS/NTU, No.1, Sec. 4, Roosevelt Rd, Taipei 10617, Taiwan}

\author[0000-0002-3468-9577]{Gaylor Wafflard-Fernandez} 
\affiliation{Univ. Grenoble Alpes, CNRS, IPAG, 38000 Grenoble, France}

\author[0000-0003-1526-7587	]{David J. Wilner} 
\affiliation{Center for Astrophysics | Harvard \& Smithsonian, Cambridge, MA 02138, USA}

\author[0000-0002-7501-9801]{Andrew J. Winter}
\affiliation{Laboratoire Lagrange, Université Côte d’Azur, CNRS, Observatoire de la Côte d’Azur, 06304 Nice, France}
\affiliation{Max-Planck Institute for Astronomy (MPIA), Königstuhl 17, 69117 Heidelberg, Germany}

\author[0000-0002-7212-2416]{Lisa W\"olfer} 
\affiliation{Department of Earth, Atmospheric, and Planetary Sciences, Massachusetts Institute of Technology, Cambridge, MA 02139, USA}

\author[0000-0001-8002-8473	]{Tomohiro C. Yoshida} 
\affiliation{National Astronomical Observatory of Japan, 2-21-1 Osawa, Mitaka, Tokyo 181-8588, Japan}
\affiliation{Department of Astronomical Science, The Graduate University for Advanced Studies, SOKENDAI, 2-21-1 Osawa, Mitaka, Tokyo 181-8588, Japan}

\author[0000-0001-9319-1296	]{Brianna Zawadzki} 
\affiliation{Department of Astronomy, Van Vleck Observatory, Wesleyan University, 96 Foss Hill Drive, Middletown, CT 06459, USA}
\affiliation{Department of Astronomy \& Astrophysics, 525 Davey Laboratory, The Pennsylvania State University, University Park, PA 16802, USA}

\begin{abstract}
Planet formation is a hugely dynamic process requiring the transport, concentration and assimilation of gas and dust to form the first planetesimals and cores. With access to extremely high spatial and spectral resolution observations at unprecedented sensitivities, it is now possible to probe the planet forming environment in detail. To this end, the exoALMA Large Program targeted fifteen large protoplanetary disks ranging between ${\sim}1\arcsec$ and ${\sim}7\arcsec$ in radius, and mapped the gas and dust distributions. $^{12}$CO J=3-2, $^{13}$CO J=3-2 and CS J=7-6 molecular emission was imaged at high angular (${\sim}~0\farcs15$) and spectral (${\sim}~100~{\rm m\,s^{-1}}$) resolution, achieving a surface brightness temperature sensitivity of ${\sim}1.5$~K over a single channel, while the 330~GHz continuum emission was imaged at 90~mas resolution and achieved a point source sensitivity of ${\sim}\,40~\mu{\rm Jy~beam^{-1}}$. These observations constitute some of the deepest observations of protoplanetary disks to date. Extensive substructure was found in all but one disk, traced by both dust continuum and molecular line emission. In addition, the molecular emission allowed for the velocity structure of the disks to be mapped with excellent precision (uncertainties on the order of $10~{\rm m\,s^{-1}}$), revealing a variety of kinematic perturbations across all sources. From this sample it is clear that, when observed in detail, all disks appear to exhibit physical and dynamical substructure indicative of on-going dynamical processing due to young, embedded planets, large-scale, (magneto-)hydrodynamical instabilities or winds.
\end{abstract}


\section{Introduction}
\label{sec:introduction}

It is now abundantly clear that the dust which settles into the planet-forming disk around a newly born star is highly structured, with numerous sources exhibiting gaps, rings, and spirals in their distributions \citep{Andrews_2020, Bae_ea_2023}. A similar substructuring in the chemical composition of the gas is observed, with varying morphologies found between both different molecules within a single source, and from system to system \citep{Oberg_ea_2023}. This ubiquity of structure points towards a highly dynamic environment, one where the gas and dust distributions are continually sculpted by \hbox{(magneto-)}hydrodynamical processes and/or a population of embedded protoplanets \citep{Bae_ea_2023, Lesur_ea_2023}.

Thus far the community has emphasized ALMA's ability to image the distribution of mm-sized grains at au-level spatial resolutions \citep{ALMA-Partnership_ea_2015, Andrews_ea_2016, Perez_ea_2019}, and for its ability to catalog the chemical complexity across the planet formation process \citep{Facchini_ea_2021, Ilee_ea_2021}. In contrast, interest in observations optimized to study gas kinematics has only recently gained momentum, but has been rapidly growing ever since \citep{Pinte_ea_2023}. Early theoretical works have explored the use of gas kinematics to detect embedded planets \citep[e.g.,][]{Perez_ea_2015}, however it was only through the claim of detection of a small sample of embedded protoplanets through their dynamical influence on the background disk that serious attention was paid to such observations \citep{Pinte_ea_2018b, Pinte_ea_2019, Teague_ea_2018a, Curone_ea_2022}. These works demonstrated that gas dynamics was an alternative to searching for the mm-continuum emission from circumplanetary disks as a planet detection method \citep[with PDS~70 being a notable exception;][]{Isella_ea_2019, Benisty_ea_2021}.

The realization that gas velocities could be constrained to a precision of less than $10~{\rm m\,s^{-1}}$ \citep{Teague_ea_2018a, Casassus_Perez_2019, Izquierdo_ea_2022} and decomposed into azimuthally-symmetric components \citep{Teague_ea_2019b, Yu_ea_2021, Izquierdo_ea_2023b}, facilitated a range of analyses that proved particularly promising for characterizing the structure of the protoplanetary disk. Notable examples include the ability to detect super-Keplerian rotation attributed to the gravitational potential of moderately sized disks leading to constraints on the disk mass \citep{Vernoesi_ea_2021, Lodato_ea_2023, Martire_ea_2024}, deep searches for kinematic evidence of the gravitational instability \citep{Paneque-Carreno_ea_2021, Speedie_ea_2024}, and the identification of the launching of a disk wind \citep{Galloway-Sprietsma_ea_2023}.


In the context of this exoALMA Special Issue, the goal of this paper is to set out the motivation, design and implementation of the exoALMA Project, providing the background for the various studies which have been conducted using the data set and is organized as follows. We describe the scientific motivation for exoALMA in Section~\ref{sec:scientific_motivation} and how this dictated the observational setup in Section~\ref{sec:program_design}. The resulting data products are discussed in Section~\ref{sec:data_products} and Section~\ref{sec:summary} provides a brief summary of the main findings from the initial analysis of the data.

\section{Scientific Motivation}
\label{sec:scientific_motivation}

With high spatial and spectral resolution and sensitive observations, there are a myriad of science goals which can be achieved. While many overlap, or are strongly connected, those targeted by the exoALMA program can be broadly categorized into the following three subsections: planet hunting, \S\ref{sec:scientific_motivation:planets}; the dynamical structure of the protoplanetary disk, \S\ref{sec:scientific_motivation:dynamics}; and the physical structure of the protoplanetary disk, \S\ref{sec:scientific_motivation:structure}.

\subsection{Embedded Protoplanets}
\label{sec:scientific_motivation:planets}

The main motivation of the project is to detect the presence of embedded protoplanets. If detected, such objects provide a unique opportunity to better understand the planet formation process by demonstrating \emph{where} planets are forming within a disk, place constraints on \emph{how quickly} they form and catalog \emph{from what} material they are forming, all vital inputs to population synthesis models \citep[e.g.,][]{Mordasini_ea_2012, Forgan_ea_2018}. These snapshots of young planetary systems are crucial for confronting the vast populations of mature exoplanetary systems discovered thus far, providing insights as to the relative importance of varied initial conditions or different formation pathways in explaining the stunning diversity in exoplanetary demographics.

Although protoplanets are expected to be sufficiently luminous at near-infrared wavelengths to be readily detected with ground-based facilities, searches at these wavelengths have had limited success. While planetary sources such as PDS~70~b and c \citep{Keppler_ea_2018, Haffert_ea_2019} and HD~169142~b \citep{Hammond_ea_2023} have been detected in multiple observations, the subtle signals of such embedded planets make identification and confirmation a hugely challenging prospect. One interpretation of this dearth of detections is that the sizable dust column from the circumstellar and circumplanetary disks result in significant levels of extinction that near infrared photons simply cannot pierce \citep{Alarcon_Bergin_2024, Choksi_Chiang_2024}. Another, more provocative, explanation is that point-source-like features are not planets at all and instead could be explained by stellar light scattered from a structured protoplanetary disk \citep{Zhou_ea_2023}, for example. Alternative detection methodologies are sorely needed.

To this end, several works have explored how such protoplanets may be detected in the sub-mm regime through their influence on their formation environment. These techniques have historically focused on localized enhancements in emission, either continuum emission, tracing a dust-rich circumplanetary disk \citep{Zhu_ea_2018}, or molecular emission, attributed to the localized heating and desorption of volatiles in the immediate vicinity of the planet \citep{Cleeves_ea_2015}. As with the near infrared searches, these approaches have had limited success, however two detections of continuum emission interpreted as circumplanetary disks have been made in PDS~70 \citep{Isella_ea_2019, Benisty_ea_2021} and well as a point source in line emission in AS~209 suggestive of a gas-rich circumplanetary disk \citep{Bae_ea_2022}. The small spatial scales of the emission and the challenges to achieve high dynamic range could be hampering such detections \citep{Andrews_ea_2021}.

A more promising avenue has been the study of how the planet may influence the structure of the gas in the protoplanetary disk and how these perturbations manifest in observations of molecular line emission \citep{Disk-Dynamics_ea_2020}. The most well studied of these interactions are the spiral shocks giving rise to spiral arms along Lindblad resonances in the midplane of the disk \citep{Rafikov_2002}, or buoyancy resonances in the atmosphere of the disk \citep{Bae_ea_2021}. While the density perturbations associated with such shocks are probably too subtle to be detected \citep{Speedie_Dong_2022}, the kinematic perturbations are expected to have velocities of up to $100~{\rm m\,s^{-1}}$ easily accessible with ALMA \citep{Bollati_ea_2021, Fasano_ea_2024}. Indeed, such perturbations have been identified and characterized in a large number of sources through localized features in the channel maps \citep[typically described as `kinks' owing to the zig-zag like morphology;][]{Pinte_ea_2018b, Pinte_ea_2019, Pinte_ea_2020, Pinte_ea_2023} or in maps of the projected velocity after subtracting a background rotation model \citep{Teague_ea_2019a, Teague_ea_2023, Garg_ea_2022}, or using `folding' of the maps to subtract azimuthally symmetric structures \citep{Huang_ea_2018, Izquierdo_ea_2021, Izquierdo_ea_2022, Izquierdo_ea_2023b, Stadler_ea_2023}.

\subsection{Dynamical Structure of Protoplanetary Disks}
\label{sec:scientific_motivation:dynamics}

In addition to interactions with embedded planets, there are a number of other mechanisms which can drive velocity perturbations within a protoplanetary disk. Of particular note are the numerous \hbox{(magneto-)}hydrodynamical instabilities which have been shown to be potentially active within a disk \citep{Lesur_ea_2023}. As these instabilities are the leading mechanism thought to facilitate the transport of angular momentum through turbulent viscosity, determining which of these instabilities dominate is a critical piece of the planet formation puzzle \citep{Manara_ea_2023} as these govern the lifetime of the disk, setting the time available to form planets.

How each instability manifests in the data is dependent on the spatial scale of the velocity perturbations it drives and can be broadly split between what we term as `macro'-scale features, which we take as structures on the ${\sim}~10$~au scale that are spatially resolved with typical observations, and `micro'-scale features, which are conversely spatially unresolved. For the former, numerical simulations have resulted in a variety of robust predictions. For example, the gravitational instability can drive large, multiple spiral arms \citep{Kratter_Lodato_2016} which could be traced through dust over-densities \citep{Perez_ea_2016}, localized heating of the gas \citep{Ilee_ea_2011}, or the velocity perturbations due to the shocks \citep{Hall_ea_2020}. Alternatively, the vertical shear instability, which has been modeled to result in large convection cells, appearing as concentric annular features in the gas velocity \citep{Barraza-Alfaro_ea_2021}, and drive the efficient vertical mixing of dust grains resulting in large dust scale heights \citep{Flock_ea_2020, Dullemond_ea_2022}. As with any search for substructure, maximizing the angular resolution of the observations will be critical in teasing out the subtle features.

On smaller scales, instabilities will manifest primarily as the broadening of molecular emission lines due to their contribution to the velocity dispersion along the line of sight \citep[e.g.,][]{Simon_ea_2015}. Searches for such broadening have aimed to use the radial and vertical variation in the level of non-thermal broadening to differentiate between different instabilities \citep{Hughes_ea_2011, Guilloteau_ea_2012, Flaherty_ea_2015}. For example, the magneto-rotational instability should exhibit a strong dependence on height, with turbulent velocities increasing towards the disk atmosphere where the gas is more highly ionized \citep{Flock_ea_2015}, in contrast to the gravitational instability where the strength of the velocity dispersions are independent of height above the midplane, but do depend on local scale height \citep{Forgan_ea_2011}. As the non-thermal velocity dispersion is expected to be small compared to the thermal velocity dispersion, high angular and spectral resolution data where the emission lines are fully resolved are key in obtaining robust constraints on the level of broadening \citep{Teague_ea_2016}.

A complementary scenario to disk evolution being governed by instabilities is the idea of a (potentially magnetized) disk wind which is able to efficiently remove material from the disk \citep{Pascucci_ea_2023}. Detection and characterization of disk winds will allow for the refinement of analytical wind models routinely used to explain the global evolution of disks \citep{Tabone_ea_2022, Somigliana_ea_2023}. Regardless of the wind launching mechanism, a disk wind is not expected to result in large velocity dispersions, but rather be dominated by large-scale flows, predominantly characterized by azimuthally-symmetric radial components \citep{Riols_Lesur_2018, Haworth_Owen_2020}. As demonstrated by \citet{Rosenfeld_ea_2014}, such radial velocities will manifest as a rotation of the projected velocity map, rather than the distinct features predicted by other instabilities. Here, both spatial and spectral resolution is necessary to decompose any wind component from the bulk Keplerian rotation of the protoplanetary disk.

\subsection{Physical Structure of Protoplanetary Disks}
\label{sec:scientific_motivation:structure}

To understand whether the conditions are suitable for the formation of planets, or the growth of a particular instability, it is necessary to gain knowledge of the gas density and temperature structure of the disk.

In the high optical depth regime, typical of millimeter $^{12}$CO, $^{13}$CO line and continuum emission, the brightness temperature acts as an accurate probe of the local gas temperature \citep{Weaver_ea_2018}. With ALMA observations now routinely resolving the vertical, as well as radial, structure of the disk, techniques like those presented in \citet{Pinte_ea_2018a} enable us to sample the two dimensional temperature structure of disks that are only moderately inclined \citep{Law_ea_2022}. Such empirical measurements are critical for confronting the temperature structures used as either input for numerical simulations, or used to interpret observations. For example, although observations of hydrogen deuteride, HD, emission have been touted as a robust probe of the disk mass, the interpretation of the emission, and resulting disk mass, is highly dependent on the assumed temperature structure \citep{McClure_ea_2016}.

The limitation of this regime, however, is that the intensity of the line core is independent of the column density, and the considerably weaker line wings must be leveraged to learn about the column density. Instead, optically thin lines such as the rarer C$^{18}$O and C$^{17}$O isotopologues would offer an opportunity to more readily trace spatial variations in the local density, however chemical processing will complicate the interpretation and limit application to the bulk gas disk \citep{Zhang_ea_2017}.

Fortuitously, the global disk density has been demonstrated to influence molecular line emission beyond its intensity, offering novel approaches to probing the disk density structure. One approach is to study the influence on the dynamics of the gas traced by the molecular emission. In addition to the gravitational potential from the host star, the radial pressure gradient and disk gravitation potential will modulate the rotation velocity of the gas, $v_{\phi}$ \citep{Rosenfeld_ea_2013}. Thus, with precise measurements of $v_{\phi}$ one can infer the background pressure profile \citep{Teague_ea_2018a, Teague_ea_2018c} and, through comparison with the dust density profiles derived from continuum observations, explore the grain-trapping potential of pressure traps \citep{Rosotti_ea_2020}. Furthermore, with assumptions about the functional form of the background temperature and density structure, modeling of the $v_{\phi}$ profile can provide constraints on the disk mass with an accuracy of ${\sim}~5\%$ \citep[at least for systems where $M_{\rm disk} \, / \, M_{*} \gtrsim 0.05$]{Andrews_ea_2024, Veronesi_ea_2024}. An alternative is to explore the influence on the profile of the line emission. For example, \citet{Yoshida_ea_2022} demonstrated that non-Gaussian line profiles are observed in regions of high density owing to the no-longer-negligible contribution of collisional broadening, providing tight constraints on the local \emph{volume} density of the disk.

Clearly, improved constraints on such fundamental properties as the gas temperature and density profiles will be hugely important for the refinement of models and simulations, both in terms of making accurate predictions and interpreting future observations. For these goals, a combination of high spatial and spectral resolution is key. The spatial resolution is necessary to appropriately sample the spatial variation in emission and minimize unresolved intensity gradients, while spectral resolution is critical for detecting the subtle variations in line profile.

\section{Observational Design}
\label{sec:program_design}

\begin{deluxetable*}{rccccCcccCCCCCC}
\tablecaption{exoALMA Sample\label{tab:sample}}
\tabletypesize{\footnotesize}
\tablehead{
\colhead{Source}  & \colhead{\bf Alias} & \colhead{R.A.} & \colhead{Dec} & \colhead{SpT} & \colhead{$d$ (pc)} & \colhead{RUWE\tablenotemark{a}} & \colhead{Refs.}
}
\decimals
\startdata          
DM~Tau              & -          & 04 33 48.733 & +18 10 09.973 & M1     & 144    & 1.49 & 1  \\
AA~Tau              & -          & 04 34 55.420 & +24 28 53.034 & M0     & 135    & 3.33 & 2  \\
LkCa~15             & -          & 04 39 17.791 & +22 21 03.390 & K5     & 157    & 1.36 & 1  \\
HD~34282            & -          & 05 16 00.477 & -09 48 35.394 & A0.5   & 309    & 1.41 & 3  \\
MWC~758             & -          & 05 30 27.529 & +25 19 57.076 & A3     & 156    & 0.99 & 4  \\
CQ~Tau              & -          & 05 35 58.467 & +24 44 54.091 & F2     & 149    & 3.70 & 4  \\
SY~Cha              & -          & 10 56 30.388 & -77 11 39.402 & K5     & 182    & 1.37 & 5  \\
PDS~66              & MP Mus     & 13 22 07.542 & -69 38 12.219 & K1     & 98     & 0.96 & 6  \\
HD~135344B          & SAO~206462 & 15 15 48.446 & -37 09 16.024 & F4     & 135    & 0.91 & 7  \\
HD~143006           & -          & 15 58 36.913 & -22 57 15.221 & G7     & 167    & 1.13 & 8  \\
RXJ1604.3-2130\,A   & J1604      & 16 04 21.642 & -21 30 29.058 & K2     & 145    & 1.75 & 8  \\
RXJ1615.3-3255      & J1615      & 16 15 20.234 & -32 55 05.098 & K5     & 156    & 1.55 & 9  \\
V4046~Sgr           & -          & 18 14 10.482 & -32 47 34.517 & K5     & 72     & 0.86 & 10 \\
RXJ1842.9-3532      & J1842      & 18 42 57.981 & -35 32 42.827 & K2     & 151    & 1.10 & 11 \\
RXJ1852.3-3700      & J1852      & 18 52 17.301 & -37 00 11.949 & K2     & 147    & 1.58 & 12
\enddata
\tablenotetext{a}{Renormalized Unit Weight Error from Gaia \citep{Gaia_DR3}.}
\tablecomments{References for spectral types: 1, \citet{Kenyon_Hartmann_1995}; 2, \citet{Herbig_1977}; 3, \citet{Pietu_ea_2003}; 4, \citet{Mannings_Sargent_2000}; 5, \citet{Frasca_ea_2015}; 6, \citet{Schutz_ea_2005}; 7, \citet{Grady_ea_2009}; 8, \citet{Luhman_Mamajek_2012}; 9, \citet{Wichmann_ea_1999}; 10, \citet{Nefs_ea_2012}; 11, \citet{Carpenter_ea_2005}; 12 \citet{Manara_ea_2014}.}
\end{deluxetable*}

Motivated by this range of science goals, it was clear that obtaining deep, spatially and spectrally resolved observations of a large sample of protoplanetary disks would offer a powerful opportunity to make significant progress in several areas. In this section, we discuss how the observations were designed to maximize the scientific potential of the data set.

\subsection{Sources}
\label{sec:program_design:sources}

When developing the source list for the exoALMA program, several aspects needed to be balanced: sources must represent new frontiers (i.e., minimize re-observing common targets), be well suited for the scientific goals, and must be readily observable throughout the cycle. The former condition essentially precluded any source which had already been observed extensively in long-baseline configurations with ALMA, such as the five sources from the MAPS Large Program \citep[AS~209, HD~163296, GM~Aur, IM~Lup and MWC~480;][]{Oberg_ea_2021}, TW~Hya \citep{Teague_ea_2019a, Teague_ea_2023}, HD~169142 \citep{Yu_ea_2021, Garg_ea_2022}, HD~97048 \citep{Pinte_ea_2019} and HD~100546 \citep{Booth_ea_2024}.

In terms of scientific applicability, four aspects were considered: brightness, radial extent, obscuration and inclination. A bright disk is necessary to image the data at a high velocity sampling (brighter sources would provide higher signal to noise ratios with narrower channel spacing), while a large disk is necessary to maximize the number of independent spatial samples across that can be measured. Verifying that there is no absorption or contamination from large-scale emission, such as a remnant envelope, would ensure that all channels would be usable in the analysis. Even if a source was large, bright and unobscured, if it is viewed at an unfavorably high inclination then many typically employed analyses would fail. Most notably, at high inclinations, $i \gtrsim 60\degr$, the near and far emission lobes characteristic of a rotating source become confused with emission arising from the top and bottom surfaces of the disk \citep[see the discussion for the highly inclined Gomez's Hamburger in][for example]{Teague_ea_2020}. While such high inclination sources offer an opportunity to study the vertical structure of the disk \citep[particularly when considering just the dust continuum;][]{Villenave_ea_2020}, the need to assume a dynamical model to `deproject' the molecular line data precludes any kinematical studies \citep{Dutrey_ea_2017, Ruiz-Rodriguez_ea_2021}.

The final criteria was the distribution of sources in right ascension to aid in scheduling observations. As Band 7 observations require typically better weather conditions than the more commonly used (at least for protoplanetary disks) Band 6, a wide spread in RA was selected to ensure that there was a high chance for the complete sample to be observed in a single cycle. In practice, this required limiting the number of sources from any given star-forming region. A discussion of why Band 7 was chosen over Band 6 can be found in the following sub-section.

These considerations led to three conditions for source selection. The sources must all have a gas extent of at least $1\arcsec$ in radius, have an inclination that falls between $5\degr$ and $60\degr$, and show no evidence of envelope emission in archival data. The total intensity of the archival $^{12}$CO emission was not used as an absolute criteria, but rather to decide between sources that otherwise appeared equally suitable given the scheduling constraints.

The resulting fifteen sources targeted with exoALMA are described in Table~\ref{tab:sample}. It is important to note that these considerations resulted in a decidedly biased sample of disks in terms of whether they represent the broader protoplanetary disk population. As demonstrated by the stellar properties listed in Table~\ref{tab:sample}, these disks tend to be around more massive and luminous stars and have a larger fraction of transition disks or azimuthal asymmetries in their dust continuum. Indeed, all but PDS~66 exhibit gap-like or cavity-like structures in their continuum emission \citep[which does show a ring in NIR scattered light;][]{Wolff_ea_2016}, as discussed in \citet{Curone_exoALMA}. While this may limit the applicability of findings to smaller, potentially more `typical' protoplanetary disks, this sample represents the best opportunity to demonstrate ALMA's ability to map gas kinematics in a protoplanetary disk, yielding the data necessary to develop the techniques necessary to extend these studies to a larger, more representative sample.

\begin{figure*}
    \centering
    \includegraphics[width=\textwidth]{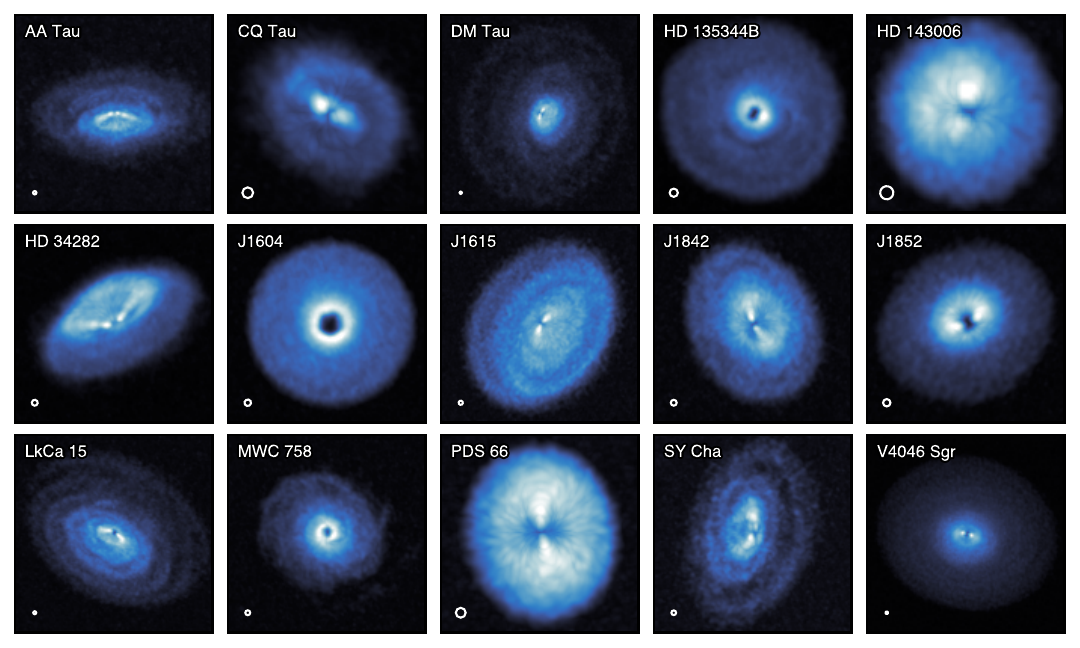}
    \caption{A summary of the exoALMA sample with all sources showing evidence of gas substructures. All images show the $^{12}$CO (3-2) peak intensity obtained with \texttt{bettermoments} \citep{Teague_Foreman-Mackey_2018} using a quadratic fit to the data and masked with the CLEAN masks. Each source has a variable field of view and the open circle in the lower left of each image denotes the $0\farcs15$ synthesized beam for reference. The color maps is individually scaled for each source to emphasize the emission morphology rather than for a quantitative comparison (which can be found in \citealt{Galloway_exoALMA}).}
    \label{fig:CO_gallery_main}
\end{figure*}

\subsection{Telescope Setup}
\label{sec:program_design:setup}

In order to achieve the aforementioned goals we opted for a relatively simple spectral setup, consisting of four spectral windows, three of which were assigned to a single molecular line, $^{12}$CO J=3-2, $^{13}$CO J=3-2 and CS J=7-6, and one for continuum emission centered at 332~GHz (0.9~mm). Optically thick CO isotopologues provide the best opportunity to trace the gas structure due to their abundance, while their large optical depth results in their observed emission arising from only a narrow layer in the disk, justifying the use of the simple `two-layer' morphological models often used for interpreting the data \citep[e.g.,][]{Teague_ea_2018a, Casassus_Perez_2019, Izquierdo_ea_2021}. With its relatively large abundance among the simple species accessible simultaneously with CO such that high SNR observations were likely to be achieved for many sources, CS was the third molecule chosen, providing access to a region close to the disk midplane \citep{Dutrey_ea_2017}. Interestingly for kinematical studies, its large molecular weight, $\mu = 44$, limits its sensitivity to thermal broadening compared to the lighter CO isotopologues ($\mu \approx 28$), making it ideal for searches for turbulence through non-thermal broadening \citep{Guilloteau_ea_2012, Teague_ea_2016, Teague_ea_2018b}.

The three line spectral windows were configured to use the maximum spectral sampling of 15~kHz, resulting in a spectral resolution of 30.5~kHz (27~${\rm m\,s^{-1}}$) when accounting for the signal processing associated with the correlator. The continuum window was centered at 332~GHz (0.9~mm) and set to frequency division mode (FDM) yielding a spectral resolution of 488.281~kHz (440~${\rm m\,s^{-1}}$) across a bandwidth of 1875~MHz. The choice to observe transitions which fall in Band 7 rather than the more commonly targeted Band 6 transitions was driven primarily by the desire to maximize the velocity resolution of the data: a frequency resolution of 30.5~kHz corresponds to a velocity resolution of 27~${\rm m\,s^{-1}}$ at 345~GHz, but only 40~${\rm m\,s^{-1}}$ at 230~GHz. This higher velocity resolution allows the FWHM of an emission line to be sampled ${\sim}10$ times allowing for a precise characterization of the line profile while also remaining sensitive to subtle deviations in the line profiles arising to due spatially unresolved structures \citep{Lenz_Ayres_1992}. Although rarely used, obtaining the data at such a fine velocity resolution provides confidence that any spectral features observed are real, rather than due to sampling or averaging effects that arise for lower spectral resolutions. Furthermore, to achieve a typical RMS sensitivity of 1.5~K, Band 7 requires a factor of {$\sim$}~3 times shorter integrations than Band 6, allowing for a substantially larger sample to be considered.

As discussed in Section~\ref{sec:scientific_motivation}, obtaining a high spatial resolution is key if features in the gas are to be accurately associated with the features observed in (typically higher spatial resolution) dust continuum observations. Guided by the images of CO isotopologue emission released by the MAPS Large Program \citep{Oberg_ea_2021}, a target spatial resolution of $0\farcs1$ (14~au at 140~pc) was chosen as a trade off between imaging quality and the general scales of features observed in the dust continuum \citep[e.g.,][]{Huang_ea_2018a}. It was found that a combination of configurations C43-3 and C43-6 would result in the desired angular resolution. A handful of sources (DM~Tau, LkCa~15, HD~34282, J1604, J1615, V4046~Sgr and AA~Tau) were known to have molecular emission extending close-to, if not beyond, the maximum recoverable scale of $4\farcs7$ and were therefore additionally observed with the ACA, extending the maximum recoverable scale to $19\farcs3$.

\subsection{Observations}

Data was collected between October 2021 and May 2023 and delivered to the collaboration after an initial quality assurance check and pipeline calibration by ALMA. \citet{Loomis_exoALMA} describes the subsequent calibration and imaging procedures conducted by the exoALMA team upon delivery of the data.

\section{The exoALMA Data Set}
\label{sec:data_products}

\begin{figure*}
    \centering
    \includegraphics[width=\textwidth]{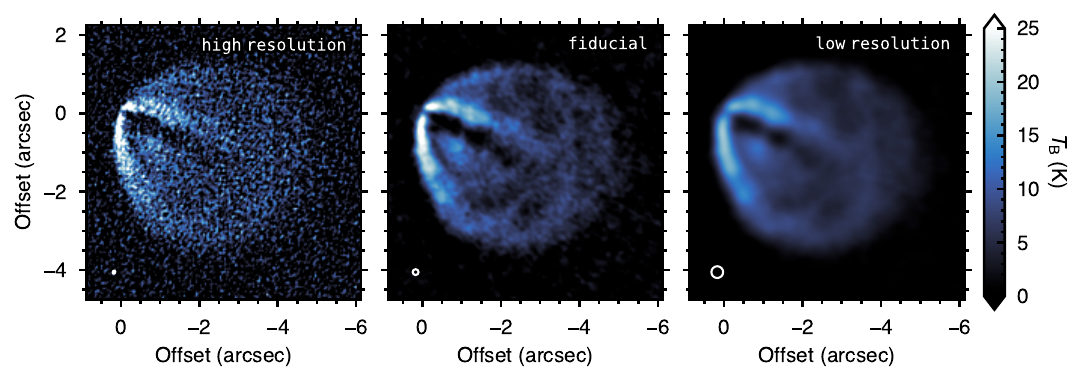}
    \caption{$^{12}$CO emission, converted to brightness temperature using the Rayleigh-Jeans approximation, from the disk around LkCa~15, demonstrating the differences between the three different imaging sets released with the exoALMA program: `high' resolution, left; `fiducial', center; and `low', right. =The spatial resolution and channel spacing for each image is, from left to right, $86~{\rm mas} \times 67~{\rm mas}$ and $200~{\rm m\,s^{-1}}$, $0\farcs15 \times 0\farcs15$ and $100~{\rm m\,s^{-1}}$ and $0\farcs3 \times 0\farcs3$ and $100~{\rm m\,s^{-1}}$. Note that the `high' resolution image sets have variable angular resolutions set by the requirement of resolving the continuum substructure.}
    \label{fig:images_sets}
\end{figure*}

\subsection{Overview}

The exoALMA Data Set comprises observations of 15 sources with emission from $^{12}$CO J=3-2, $^{13}$CO J=3-2, CS J=7-6 and 332~GHz continuum. In addition to strong detections of the continuum emission, all molecules were robustly detected in every source and no cloud contamination was found for any of the lines. 

Figure~\ref{fig:CO_gallery_main} presents an overview of the continuum subtracted $^{12}$CO peak intensity of the exoALMA sample using the `fiducial' imaging set, described below, highlighting the diversity in morphologies and structures observed. It is clear from this figure alone that these disks have a gas structure which is highly structured, perhaps unsurprising given the apparent ubiquity of structure observed in the dust continuum \citep{Andrews_2020, Curone_exoALMA}. Similar galleries for the $^{13}$CO and CS emission can be found in Appendix~\ref{app:galleries}.

\subsection{Data Products}

\begin{figure*}
    \centering
    \includegraphics[width=\textwidth]{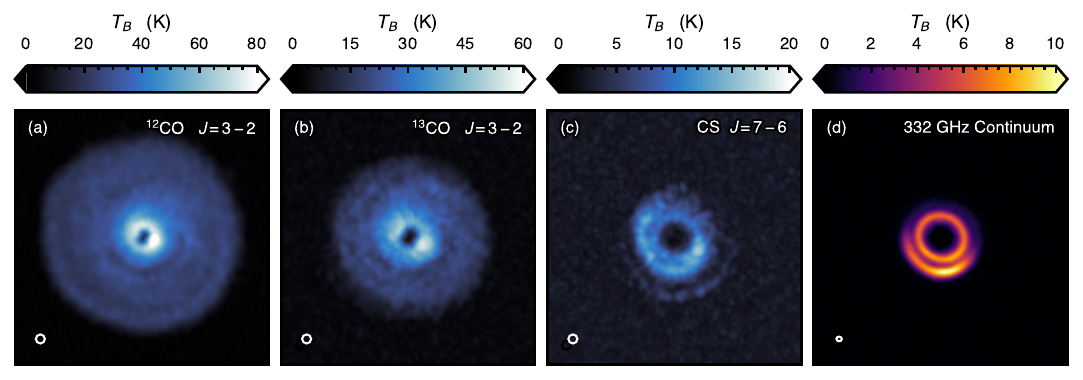}
    \caption{Example of the peak intensities derived from the `fiducial set' of images for HD~135344B. All panels share a $4\arcsec\! \times 4\arcsec$ field of view centered on the star. The spatial resolution of molecular emission images is $0\farcs15$ and the channel spacing is 100~${\rm m\,s^{-1}}$ for both $^{12}$CO and $^{13}$CO and 200~${\rm m\,s^{-1}}$ for CS. The continuum image has a synthesized beam of $90~{\rm mas} \times 76~{\rm mas}$ with a position angle of $85\fdg0$.}
    \label{fig:HD13344_gallery}
\end{figure*}

To accurately quantify and characterize the level and complexity of gaseous substructures hosted by these disks, a range of data products were produced and analyzed as part of the collaboration using a range of techniques described in \citet{Izquierdo_exoALMA}. Prior to the application of these techniques all pieces of software used were bench-marked to ensure that a) all predictions and retrievals are accurate and b) there is consistency among the tools adopted within the exoALMA collaboration \citep{Bae_exoALMA}.

\subsubsection{Image Sets}
\label{sec:data_products:image_sets}

A number of different image sets for the line and continuum emission were produced to satisfy the range of different criteria required by the various projects: a `fiducial' set, a `high resolution' set and a `low resolution' set. The `fiducial' set was designed to be appropriate for the broadest range of scientific questions with moderate angular and spectral resolutions, the `high resolution' sets focused on maximizing angular resolution while sacrificing spectral sampling or sensitivity, and the `low resolution' enhanced the surface brightness sensitivity of the images. A comparison of channel maps from each of these sets is shown in Fig.~\ref{fig:images_sets} for the case of LkCa~15, and the implementation and results of the imaging process discussed in \citet{Loomis_exoALMA} and summarized in their Tables~3, 4 and 5.

\paragraph{Fiducial Images}

The fiducial images used a variable robust value with a source-specific $uv$-taper to obtain a circular $0\farcs15$ synthesized beam and were imaged with a channel spacing of $100~{\rm m\,s^{-1}}$ for $^{12}$CO and $^{13}$CO, and $200~{\rm m\,s^{-1}}$ for CS. The circularized beam, rather than the elliptical beam that is obtained by default, was deemed to introduce less confusion in the interpretation of features in image-plane analyses. The $100~{\rm m\,s^{-1}}$ velocity spacing was found to provide a good trade-off between spectrally resolving the line and maintaining a strong signal in a single channel for imaging. The considerably weaker emission of CS compared to the CO isotopologues required a coarser velocity sampling of $200~{\rm m\,s^{-1}}$ to conserve the per-channel imaging quality. Figure~\ref{fig:HD13344_gallery} shows an example of the `fiducial' image set for HD~135344B, displaying maps of the peak intensity of the molecular emission, along with a continuum emission. For the case of the continuum images, a default robust value of -0.5 was chosen which resulted in a typical beam size of ${\sim}~90$~mas. No tapering was applied resulting in a non-circular beam, unlike the line data. An example of a fiducial continuum image is shown in Figure~\ref{fig:HD13344_gallery}d. A thorough discussion of the continuum imaging process can be found in \citet{Curone_exoALMA}.

\paragraph{High Resolution Images}

A major goal of the exoALMA program is to compare the structure found in the gas with the dust distribution \citep[e.g.,][]{Stadler_exoALMA, Yoshida_exoALMA}. However, the $0\farcs15$ resolution of the fiducial images, ${\sim}20$~au at 140~pc, is insufficient to resolve variations across the observed dust substructures \citep{Curone_exoALMA}. This motivated the `high resolution' imaging set, where each source was imaged with a robust value that yields a synthesized beam that is ${\sim}~0.5$ times smaller than the average dust substructure size for that source, resulting in beams from 65~mas for HD~34282, up to the fiducial 150~mas. For these images, the channel spacing was reduced for most sources as it was found that a coarser sampling of $200~{\rm m\,s^{-1}}$ was necessary to maintain a robust detection of the emission to accurately trace the large velocity gradients found in the inner regions of the disk, except for AA~Tau and J1604, where the $100~{\rm m\,s^{-1}}$ channel spacing was able to be maintained. At these higher angular resolutions, the sensitives dropped by a factor of ${\sim}~5$ relative to the `fiducial set', with typical RMS brightness temperatures of 7.5~K. Table~4 in \citet{Loomis_exoALMA} details the angular resolution and channel spacing adopted for each source along with the resulting noise level.

\paragraph{High Surface Brightness Sensitivity Images}

A similar limitation was faced in the outer regions of the disk with the fiducial images: weaker, more extended emission was less well reconstructed resulting in a poor determination of the properties of the edge of the disk. This prompted the generation of `low resolution' images which were identical to the fiducial images, except with a circularized $0\farcs3$ synthesized beam. While losing information on smaller spatial scales compared to the other two image sets, this set provided a considerably better representation of the large-scale emission distribution, particularly for the larger, more diffuse disks such as DM~Tau and LkCa~15, where determination of the emission surface was a challenge \citep{Galloway_exoALMA}. These images typically had RMS brightness temperatures around 0.25~K in $100~{\rm m\,s^{-1}}$ channels. 

\paragraph{Regularized Maximum Likelihood Imaging}

In addition to the standard CLEAN-based imaging, \citet{Zawadzki_exoALMA} explores the use of Regularized Maximum Likelihood (RML) imaging which uses a different set of assumptions when creating the image compared to CLEAN. Specifically, in RML imaging, the model is the image itself and the user provides information about how the pixels values can vary spatially and in relation to nearby pixels. Conversely, CLEAN builds an image through the summation of multiple Gaussian components. This difference in image reconstruction was used as a test of the robustness of the features found in the images of the molecular emission for several sources \citep[AA~Tau, HD~135344B, J1604, J1615, J1842, LkCa~15 and SY~Cha;][]{Pinte_exoALMA}, and the sensitivity of emission surface and gas temperature extraction for sources with large, diffuse emission.

\begin{deluxetable*}{rCCCCCCCc}
\tablecaption{exoALMA Derived Source Properties\label{tab:source_properties}}
\tabletypesize{\footnotesize}
\tablehead{
\colhead{Source} & \colhead{$M_*$} & \colhead{$i$\tablenotemark{a}} & \colhead{PA\tablenotemark{b}} & \colhead{$F_{^{12}{\rm CO}}$} & \colhead{$F_{^{13}{\rm CO}}$} & \colhead{$F_{\rm CS}$} & \colhead{$F_{\rm cont}$} & \colhead{\bf Continuum} \\
\colhead{} & \colhead{\footnotesize ($M_{\odot}$)} & \colhead{\footnotesize ($\degr$)} & \colhead{\footnotesize ($\degr$)} & \colhead{\footnotesize (${\rm Jy~km\,s^{-1}}$)} & \colhead{\footnotesize (${\rm Jy~km\,s^{-1}}$)} & \colhead{\footnotesize (${\rm Jy~km\,s^{-1}}$)} & \colhead{\footnotesize (${\rm mJy}$)} & \colhead{\bf Substructure\tablenotemark{c}}
}
\decimals
\startdata          
DM~Tau              & 0.45              &  38.7  & 335.7     & 33.37 \pm 0.50    & \phn8.99 \pm 0.31     & 0.70 \pm 0.06     & 226.5 \pm 0.6 & R, A, I \\
AA~Tau              & 0.79              & -58.7  & 272.7     & 22.77 \pm 0.38    & \phn4.82 \pm 0.21     & 1.82 \pm 0.17     & 189.4 \pm 0.3 & R, I, W \\
LkCa~15             & 1.14              &  50.3  & 62.1      & 33.83 \pm 0.40    & \phn9.83 \pm 0.39     & 1.59 \pm 0.12     & 407.1 \pm 0.4 & R, C \\
HD~34282            & 1.61              & -58.3  & 117.4     & 17.82 \pm 0.06    & \phn6.03 \pm 0.06     & 0.66 \pm 0.03     & 343.4 \pm 0.3 & R, A \\
MWC~758             & 1.40              &  19.4  & 240.3     & 21.02 \pm 0.23    & \phn5.46 \pm 0.19     & 0.50 \pm 0.20     & 214.5 \pm 0.2 & R, A, I \\
CQ~Tau              & 1.40              & -36.3  & 235.1     & \phn9.89 \pm 0.11 & \phn3.33 \pm 0.13     & 0.64 \pm 0.05     & 431.9 \pm 0.3 & R, S, C \\
SY~Cha              & 0.77              & -52.4  & 346.2     & 17.39 \pm 0.32    & \phn3.57 \pm 0.27     & 0.85 \pm 0.08     & 158.4 \pm 0.5 & R, A, I \\
PDS~66              & 1.28              & -31.9  & 189.0     & 11.00 \pm 0.10    & \phn2.72 \pm 0.10     & 1.07 \pm 0.07     & 336.1 \pm 0.2 & \\
HD~135344B          & 1.61              & -16.1  & 242.9     & 18.00 \pm 0.13    & \phn6.19 \pm 0.14     & 0.52 \pm 0.07     & 424.7 \pm 0.2 & R, A, C \\
HD~143006           & 1.56              & -16.9  & 168.9     & \phn6.42 \pm 0.08 & \phn1.84 \pm 0.08     & 0.40 \pm 0.04     & 155.5 \pm 0.2 & R, A, I, W \\
RXJ1604.3-2130\,A   & 1.29              &  6.0   & 258.1     & 18.29 \pm 0.45    & \phn7.01 \pm 0.55     & 1.26 \pm 0.18     & 198.4 \pm 0.3 & R, C \\
RXJ1615.3-3255      & 1.14              &  46.5  & 325.4     & 30.51 \pm 0.30    & 10.28 \pm 0.22        & 0.96 \pm 0.12     & 386.0 \pm 0.7 & R, I \\
V4046~Sgr           & 1.73$^{\rm d}$    & -34.1  & 255.7     & 59.89 \pm 0.42    & 17.05 \pm 0.24        & 3.01 \pm 0.11     & 668.4 \pm 1.0 & R, I \\
RXJ1842.9-3532      & 1.07              &  39.4  & 205.9     & 14.75 \pm 0.16    & \phn3.11 \pm 0.13     & 0.61 \pm 0.05     & 141.5 \pm 0.2 & R, C \\
RXJ1852.3-3700      & 1.03              & -32.7  & 117.1     & 10.83 \pm 0.13    & \phn3.69 \pm 0.11     & 1.66 \pm 0.05     & 150.9 \pm 0.1 & R, C
\enddata
\tablecomments{$M_*$, $i$ and PA derived from \texttt{discminer} analyses of the $^{12}$CO emission and are described in \citep{Izquierdo_exoALMA}. Typical \texttt{discminer} uncertainties are around $0.1\%$ for these parameters, however \citet{Hilder_exoALMA} suggests spatial correlations can increase these by a factor of ${\sim}\!10$. The fluxes, $F_{^{12}{\rm CO}}$, $F_{^{13}{\rm CO}}$ and $F_{\rm CS}$, were calculated by integrating zeorth moment maps in a circular aperture with a radius of $1.2 \times r_{\rm out}$ where $r_{\rm out}$ was taken from \citet{Galloway_exoALMA}. $F_{\rm cont}$ is calculated from the fiducial continuum images, as described in \citep{Curone_exoALMA}.}
\tablenotetext{a}{As discussed in \citet{Izquierdo_exoALMA}, the negative sign is used to assign whether the northern edge of the disk is tilted towards or away from us. For comparisons with continuum data, taking the absolute value is appropriate.}
\tablenotetext{b}{Defined as the angle between north and the red-shifted major axis of the disk in a counter-clockwise direction.}
\tablenotetext{c}{Types of substructure include rings, R; azimuthal asymmetries, A; spirals, S; inner disk, I; cavity, C; and warps, W, based on \citet{Curone_exoALMA}.}
\tablenotetext{d}{Combined dynamical mass of the spectroscopic binary.}
\end{deluxetable*}
\vspace{-2em}

\paragraph{Channel Spacing}

\begin{figure*}
    \centering
    \includegraphics[width=\textwidth]{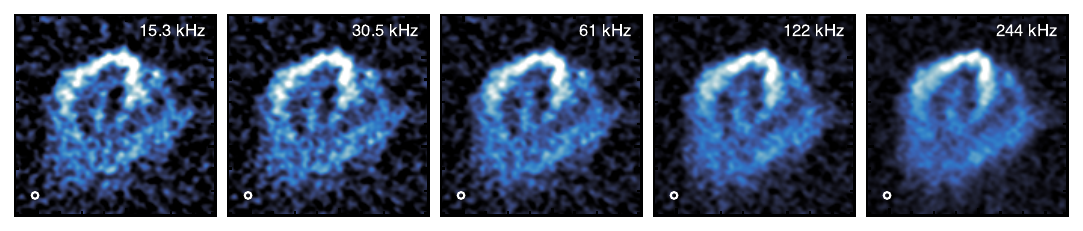}
    \includegraphics[width=\textwidth]{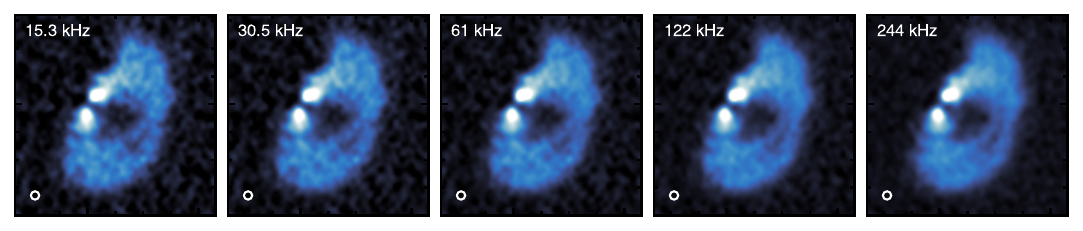}
    \caption{Channel maps of $^{12}$CO emission from SY~Cha, top, and HD~135344B, bottom, showing the effect of different channel widths. The channel widths, labeled in the top of each panel, correspond to, at the rest frequency of $^{12}$CO, 13.3, 26.4, 52.9, 105.8 and 211.5~${\rm m\,s^{-1}}$. Each channel was CLEANed down to 3$\times$ the RMS measured in a line-free region of the image. Note that while the background noise is reduced with coarser channel widths, structures in the emission morphology can be washed out. Also note that the finest channel sampling of 15.3~kHz, that native channel spacing, is actually \emph{below} the spectral resolution of the data so would suffer from considerable covariance with adjacent channels \citep{Loomis_ea_2018}.}
    \label{fig:channel_spacing_example}
\end{figure*}

As shown in Fig.~\ref{fig:channel_spacing_example}, the data could be imaged with narrower or broader channel spacing as long as this is no finer than the spectral resolution of the data ($30.5~{\rm kHz}~{\approx}~27~{\rm m\,s^{-1}}$). However, the quality of the non-Keplerian features is impacted depending on what channel spacing is chosen. With broad channel spacing, kinematic features are poorly resolved spectrally (essentially smoothing them out in the channel) and thus are hard to identify in the images. For narrow channel spacing, the features are spectrally well-resolved, but there is insufficient signal in a single channel for a robust image reconstruction, often leading to confusion as to what is a real feature. In the `fiducial' imaging sets generated for exoALMA, the finest channel spacing adopted was $100~{\rm m\,s^{-1}}$ for $^{12}$CO and $^{13}$CO emission, and $200~{\rm m\,s^{-1}}$ for CS emission as these were found to provide a good balance between excellent imaging quality and spectrally resolving the features of interest. The raw data and calibrated measurement sets from the project, however, supports the imaging of data at a finer channel spacing if one so desires.

\subsubsection{General System Properties}

For all sources, general system properties were derived following analyses of the continuum and line emission described in detail in \citet{Curone_exoALMA}, \citet{Izquierdo_exoALMA} and \citet{Galloway_exoALMA}, respectively, and provided in Table~\ref{tab:source_properties}. These include the orientation of the disk on the sky, parameterized by the inclination, $i$, and position angle, PA, measured from north to the red-shifted major axis in an anti-clockwise direction. These were used in conjunction with Gaia-derived distances for the sources \citep{Gaia, Gaia_DR3}, summarized in Table~\ref{tab:sample}, to derive a dynamical stellar mass of the host star using the \texttt{discminer} modeling approach described in \citet{Izquierdo_exoALMA}. This kinematically derived stellar mass is the stellar mass which best describes the observed rotation of the gas assuming a purely Keplerian profile and neglecting the effects of vertical stratification, radial pressure gradients or self-gravity which are discussed in \citet{Stadler_exoALMA} and \citet{Longarini_exoALMA}. Table~\ref{tab:source_properties} also includes a summary of the sorts of substructures detected in the continuum, with \citet{Curone_exoALMA} providing more details.

Integrated intensities of each of the three emission lines and the integrated flux density of the continuum emission for each source were also calculated. For the calculation of the fluxes, the continuum subtracted `fiducial' image set was used along with a circular aperture with a radius of $1.2 \times r_{\rm out}$ where $r_{\rm out}$ was taken from \citet{Galloway_exoALMA}. The flux density of the continuum also used the `fiducial' continuum images and integrated the emission within a source-specific mask. The reader is referred to \citet{Curone_exoALMA} for more details. For all four measurements, the ${\sim}10\%$ uncertainties associated with flux calibration are significantly larger than the statistical uncertainties due to noise in the data.

\subsubsection{Moment Maps}

The line emission for each source was summarized in different moment maps: an integrated intensity map (`zeroth moment' map), a peak intensity map, a line center map, and a line width map. As discussed in \citet{Teague_Foreman-Mackey_2018}, there are multiple approaches to deriving the line center (and analogously, the line width) which involve different assumptions about the intrinsic line profile and orientation of the disk on the sky. \citet{Izquierdo_exoALMA} discusses the various approaches taken by the exoALMA collaboration, including a decomposition of the emission arising from the front and back sides of the disk \citep[e.g.,][]{de-Gregorio-Monsalvo_ea_2013, Rosenfeld_ea_2013}, to calculate these maps.

\subsubsection{Vertical and Radial Emission Structure}

\begin{figure*}
    \centering
    \includegraphics[width=\textwidth]{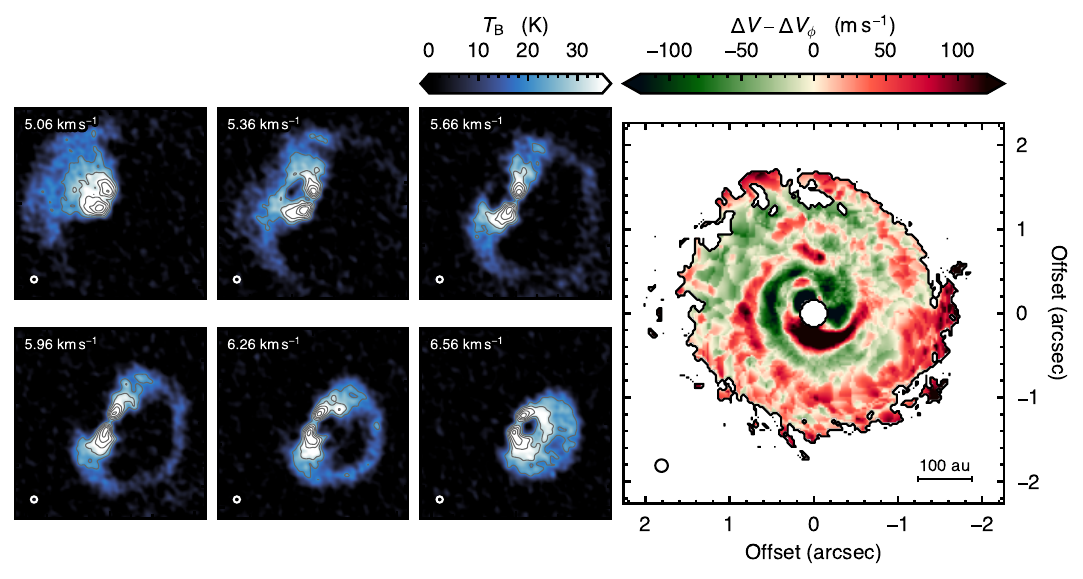}
    \caption{The high sensitivity of the exoALMA data reveals structures and subtle perturbations previously undetectable. As an example, the channel maps show $^{12}$CO emission from the disk around MWC~758 which exhibits significant non-Keplerian components, such as the large arc in the south-west of the disk. \citet{Pinte_exoALMA} leverage this sensitivity to look for indications of embedded planets in the exoALMA sources. Similarly, the right panel shows the variations in line width of the $^{12}$CO emission after subtracting an azimuthally averaged value, revealing a striking spiral morphology. \citet{Izquierdo_exoALMA} describes how such analyses were conducted for the exoALMA sample and how they were characterized. All panels share the same field of view and the $0\farcs15$ beam is shown in the lower left corner of each.}
    \label{fig:MWC758_channels}
\end{figure*}

The moment maps described above can be further collapsed into one dimensional radial profiles through an azimuthal average. \citet{Curone_exoALMA} and \citet{Galloway_exoALMA} discuss the application to the continuum and the molecular line peak intensity, respectively, and catalog the features, such as gaps and rings, found in the radial profiles, following the procedure described in \citet{Huang_ea_2018a} and \citet{Law_ea_2021a}. Notably, most sources were found to exhibit structure in their radial emission profiles aside from PDS~66, consistent with previous findings \citep{Ribas_ea_2023}.

For sources with an inclination greater than $20\degr$, a majority of the sample, the emission heights of the three molecular lines were determined using the method presented in \citet{Pinte_ea_2018a} as implemented in \texttt{disksurf} \citep{disksurf} and described in detail in \citet{Galloway_exoALMA}, and using \texttt{discminer} \citep{Izquierdo_ea_2021} and described in \citet{Izquierdo_exoALMA}. These two approaches are complementary as \texttt{disksurf} recovers a non-parametric emission surface after adopting general source parameters, while \texttt{discminer} fits a parameterized disk model to the data, including a smooth emission surface, while allowing the general source parameters to freely vary. A more thorough comparison of these methods can be found in \citet{Galloway_exoALMA}. The three molecules trace distinct vertical regions within the disk and provide probes of the physical and dynamical conditions at those heights, enabling studies of the vertical dependence of physical processes within the disk. As with the substructure identification in the radial intensity profiles, \citet{Galloway_exoALMA} characterizes variations in the emission surface and discusses (anti-)correlations with the substructures found in the dust continuum.

\subsubsection{Temperature Structures}

Assuming that the $^{12}$CO and $^{13}$CO emission is optically thick, the brightness temperature as a function of radius and height can be used to derive empirical two-dimensional temperature structures for all the sources \citep[e.g.,][]{Law_ea_2021b}. \citet{Galloway_exoALMA} uses these profiles, derived from the `fiducial' images, to fit standard parametric forms for the two-dimensional temperature structure \citep{Dartois_ea_2003} to each of the sources.

\subsubsection{Velocity Profiles}

With constraints on the height above the midplane that the molecular emission arise from in hand, the emission morphology could be correctly deprojected and rotation profiles derived for each molecular line and each source \citep[correctly accounting for the projection effects of a vertically-extended disk is crucial in deriving accurate rotation curves, e.g.,][]{Andrews_ea_2024}. \citet{Stadler_exoALMA} describes the process of extracting the rotation profiles and identifies subtle radial variations which can be attributed to underlying pressure gradient variations \citep{Teague_ea_2018a, Teague_ea_2018c, Rosotti_ea_2020}. A disk-wide analysis of the profiles is described in \citet{Longarini_exoALMA} which allows for robust determinations a dynamical disk mass for 10 sources \citep[e.g.,][]{Vernoesi_ea_2021, Lodato_ea_2023}. Dynamical disk masses could not be measured for sources where strong azimuthal asymmetries in the disk emission prevented reliable $v_{\phi}$ measurements (MWC~758 and CQ~Tau), or sources viewed at low inclinations such that the vertical emission height of the molecule could not be determined \citep[HD~135344B, HD~143006 and RXJ1604.3-2130 A;][]{Longarini_exoALMA}.

\begin{figure*}
    \centering
    \includegraphics[]{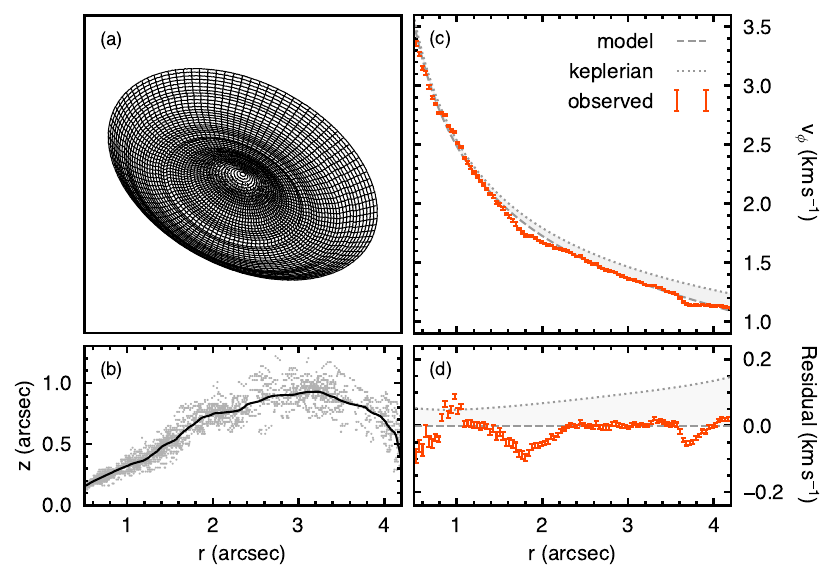}
    \caption{The high angular and spectral resolution of the exoALMA data enable a detailed exploration of the 3D physical and dynamical structure of protoplanetary disks. Panel (a) shows the projection of the inferred emission surface of the $^{12}$CO emission from the disk around LkCa~15 shown in panel (b). As described in \citet{Galloway_exoALMA}, the gray points are individual measurements while the black line shows a moving average with a window of $0\farcs15$. This deprojection allows for a precise $v_{\phi}$ profile to be extracted, as shown by the orange points in panel (c), which reveals a significant departure from purely Keplerian rotation, as shown by the dotted gray line \citep{Stadler_exoALMA}. The global profile can be well recovered with a model, shown by the gray line, that includes a radial pressure gradient and the inclusion of the self-gravity of the disk \citep{Longarini_exoALMA}. The localized variations, most readily discernible in the residuals shown in panel (d), can be associated with localized pressure variations within the disk \citep{Stadler_exoALMA}.}
    \label{fig:LkCa15_velocity}
\end{figure*}

As is clear from Fig.~\ref{fig:CO_gallery_main}, all sources exhibit some level of azimuthal structure which will bias the recovered rotation profiles as the extraction methods assume that any azimuthal variation is due to azimuth-dependent projection effects rather than intrinsic azimuthal variations \citep{Pinte_ea_2023}. However, as discussed in \citet{Izquierdo_exoALMA}, testing has demonstrated these biases are small in all but the most extreme of cases and can thus be used with confidence to study the global structure of the protoplanetary disk.

A calculation and analysis of the azimuthally averaged radial profiles of the radial and vertical velocities will be discussed in a forthcoming series of papers.

\subsection{Data Access}

All of the data products will be publicly available through both the ALMA archive as products associated with project ID \texttt{2021.1.01123.L}, and through the exoALMA website\footnote{www.exoalma.com}.

\section{Summary of exoALMA Results}
\label{sec:summary}

\begin{figure*}
    \centering
    \includegraphics[width=\textwidth]{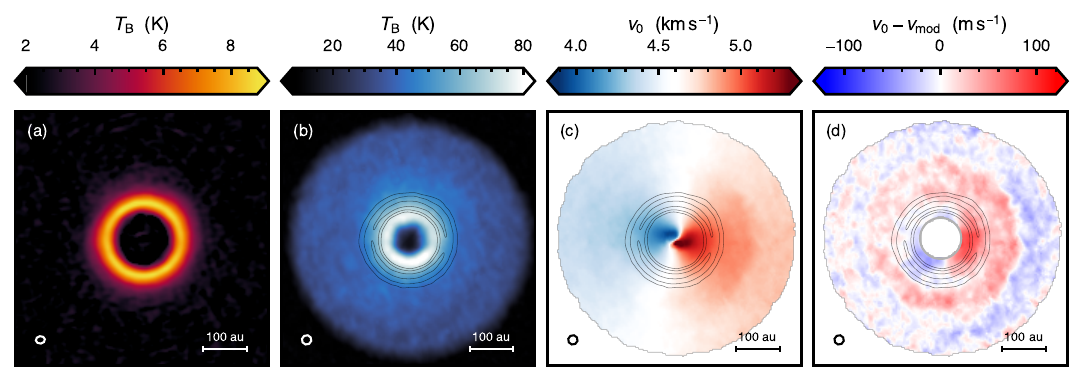}
    \caption{Combined analyses of the gas structure and dynamics with the dust substructure can provide insights as to the physical processes sculpting the disk. These four panels display emission from the disk around RXJ1604.3-2130 A with (a) showing the dust continuum reported in \citet{Curone_exoALMA}, the $^{12}$CO peak intensity in (b) and the inferred projected velocity, $v_0$, in (c), with \citet{Izquierdo_exoALMA} describing the techniques used to calculate these. Panel (d) shows the residual velocity structure after subtracting a purely Keplerian model, revealing substantial deviations which can be associated with pressure variations \citep{Stadler_exoALMA, Yoshida_exoALMA}.}
    \label{fig:J1604_moments}
\end{figure*}

The exoALMA Project surveyed fifteen protoplanetary disks and obtained high spatial (${\sim}~0\farcs15$) and spectral (${\sim}~100~{\rm m\,s^{-1}}$) resolution observations of $^{12}$CO J=3-2, $^{13}$CO J=3-2, CS J=7-6 and 332~GHz continuum emission enabling a comprehensive mapping of the physical and dynamical structure of the planet forming environment. These observations definitively show that protoplanetary disks harbor extensive substructure in their gas distributions, evident in the density, temperature and kinematics of the disk.

This observing program has provided a unique look at the planet-forming environment; one that has challenged the conventional understanding of disk structure, dynamics and contemporary analysis techniques. The complexity of these observations catalyzed a comprehensive assessment of the commonly adopted imaging techniques \citep{Loomis_exoALMA, Zawadzki_exoALMA}, numerical methods \citep{Bae_exoALMA} and analysis methodologies \citep{Hilder_exoALMA, Izquierdo_exoALMA} to ensure information was robustly and reproducibly extracted from the data. An initial, and by no means exhaustive, analysis of the data using these techniques, described in the articles published alongside this one, demonstrates the huge potential of this dataset. A summary of the main findings from the first wave of exoALMA papers is discussed below, while results more broadly related to the 2D velocity structure will be the focus of future publications. 

\vspace{0.2em}
\noindent\emph{1. High resolution observations revealed that protoplanetary disks exhibit extensive substructure in molecular line emission.} Consistent with previous studies of protoplanetary disks, all exoALMA sources exhibit substructure in their dust continuum \citep{Curone_exoALMA}. While predominantly annular in nature, several sources host localized clumps and azimuthal asymmetries. Strikingly, all sources also displayed extensive structure in their molecular emission morphology, most readily seen in the $^{12}$CO emission, but also in the $^{13}$CO and CS emission for several brighter sources \citep{Izquierdo_exoALMA, Pinte_exoALMA, Zawadzki_exoALMA}, as presented in Fig.~\ref{fig:MWC758_channels}. Variations in the optically thick $^{12}$CO emission may be attributed to localized changes in gas temperature, surface density or non-Keplerian motion, \citep{Galloway_exoALMA, Stadler_exoALMA}, while similar variations found in the line wings, or in the optically thinner $^{13}$CO and CS emission, pointed towards variations in gas density \citep{Yoshida_exoALMA}. Observations of the quality obtained by the exoALMA project unambiguously demonstrate the ubiquity of localized structure outside ${\sim}15$~au, the bulk of the disk for these sources, ideal for revealing the processes which sculpt the disk such as planet-disk interactions, (magneto-)hydrodynamical instabilities or stellar fly-bys.

\vspace{0.2em}
\noindent\emph{2. Characterization of the vertical structure of moderately inclined protoplanetary disk yielded novel insights to the physical structure of those disks.} \citet{Galloway_exoALMA} and \citet{Izquierdo_exoALMA} measured the emitting surfaces for all molecules for each source with a favorable inclination, finding the vertical chemical stratification expected for the $^{12}$CO, $^{13}$CO and CS emission. The spatially resolved vertical structure, as demonstrated in Fig.~\ref{fig:LkCa15_velocity}, enabled the derivation of empirical 2D temperature distributions \citep{Galloway_exoALMA}, as well as constraints on the CO column density and gas surface density profiles \citep{Rosotti_exoALMA}. Critically, the improved angular resolution and sensitivity revealed many structures in the emission surfaces previously undetected which indicate substantial variations in the density structure of disks. Furthermore, with a large sample of emission surfaces to hand, a linear relationship between the average emission height, $z/r$, and the disk mass, was observed \citep{Galloway_exoALMA}.

\vspace{0.2em}
\noindent\emph{3. The extraction of extremely precise rotation curves that achieve a precision of down to ${\sim}10~{\rm m\,s^{-1}}$} revealed kinematic perturbations, up to 15\% relative to the background rotation, indicative of gas surface density variations correlated with dust substructures. Figure~\ref{fig:LkCa15_velocity}c and d demonstrate the quality of the $v_{\phi}$ profiles that were able to be derived for LkCa~15 using the exoALMA data. Interpreting these variations as pressure variations, \citet{Stadler_exoALMA} demonstrated that pressure maxima and minima are coincident with the spatially resolved continuum rings and gaps, highly suggestive of grain trapping. These findings are strengthened by the detection of the pressure broadening of $^{12}$CO emission at the location of a bright continuum ring in the disk of RXJ1604.3-2130 A, suggesting a dust trapping scenario \citep{Yoshida_exoALMA}. For most cases where the dust structures can be spatially resolved by these data, the recovered velocity profiles convincingly show that the grains are dynamically trapped in pressure maxima.

\vspace{0.2em}
\noindent\emph{4. Mapping the influence of the gravitational potential of the protoplanetary disk on the rotation profiles facilitates powerful constraints on the gas surface density and enable a calibration of chemistry-based mass measurements.} \citet{Longarini_exoALMA} was able to self-consistently model the gas rotational velocity profiles for ten sources to place tight constraints on the disk mass and gas surface density profile, finding that all sources should be stable against the gravitational instability. Figure~\ref{fig:LkCa15_velocity}c and d compare a purely Keplerian model with a dotted line, and the best fit model for LkCa~15 from \citet{Longarini_exoALMA} as a dashed gray line, clearly demonstrating a significant departure from Keplerian rotation due to the self gravity of the disk and the radial pressure gradient. In parallel, \citet{Trapman_exoALMA} used complementary observations of N$_2$H$^+$ to calibrate commonly used chemistry-based approaches to disk mass measurement, finding broad agreement --- a factor of ${\sim}3$ --- between the two methods.

\vspace{0.2em}
\noindent\emph{5. A combined analysis of dust and gas substructures allows for direct comparisons with numerical simulations to identify the processes which are driving the observed features.} Figure~\ref{fig:J1604_moments} provides an example of such a comparison for RXJ1604.3-2130 A, contrasting the continuum and $^{12}$CO emission morphology with the derived gas velocity structure. \citet{Barraza_exoALMA} explored the ability of exoALMA quality observations in differentiating between different (magneto-)hydrodynamical instabilities based on the morphology of the temperature, density and kinematic perturbations they drive. \citet{Gardener_exoALMA} conducted an in depth exploration of the LkCa~15 disk and the possibility of embedded planets driving the structures observed and \citet{Wölfer_exoALMA} scrutinized the dynamics associated with azimuthal structures in the continuum believed to be vortices. \citet{Pinte_exoALMA} discussed these localized perturbations in the context of possible embedded planets.

\vspace{0.2cm}
These works demonstrate that investing in observations that push the spatial and spectral resolution of ALMA while achieving the sensitivity necessary to map out molecular line emission will facilitate significant leaps in our understanding of the planet formation environment. Although the generalization of these findings to the global population of protoplanetary disks may be hampered by the sample selection, this program has unambiguously demonstrated the utility of deep, long-baseline observations of protoplanetary disk in probing the physical conditions of planet formation. The extensive data release will hopefully seed many new projects that generate new insights into how the planet formation process proceeds.

\section{Acknowledgments}

We thank the anonymous referee for their comments which greatly improved the quality of this paper. This paper makes use of the following ALMA data: ADS/JAO.ALMA\#2021.1.01123.L. ALMA is a partnership of ESO (representing its member states), NSF (USA) and NINS (Japan), together with NRC (Canada), MOST and ASIAA (Taiwan), and KASI (Republic of Korea), in cooperation with the Republic of Chile. The Joint ALMA Observatory is operated by ESO, AUI/NRAO and NAOJ. The National Radio Astronomy Observatory is a facility of the National Science Foundation operated under cooperative agreement by Associated Universities, Inc. We thank the North American ALMA Science Center (NAASC) for their generous support including providing computing facilities and financial support for student attendance at workshops and publications. JB acknowledges support from NASA XRP grant No. 80NSSC23K1312. MB, DF, JS have received funding from the European Research Council (ERC) under the European Union’s Horizon 2020 research and innovation programme (PROTOPLANETS, grant agreement No. 101002188). Computations by JS have been performed on the `Mesocentre SIGAMM' machine, hosted by Observatoire de la Cote d’Azur. PC and LT acknowledge support by the Italian Ministero dell'Istruzione, Universit\`a e Ricerca through the grant Progetti Premiali 2012 – iALMA (CUP C52I13000140001) and by the ANID BASAL project FB210003. NC has received funding from the European Research Council (ERC) under the European Union Horizon Europe research and innovation program (grant agreement No. 101042275, project Stellar-MADE). SF is funded by the European Union (ERC, UNVEIL, 101076613), and acknowledges financial contribution from PRIN-MUR 2022YP5ACE. MF is supported by a Grant-in-Aid from the Japan Society for the Promotion of Science (KAKENHI: No. JP22H01274). CH acknowledges support from NSF AAG grant No. 2407679. IH, CH and TH are supported by an Australian Government Research Training Program (RTP) Scholarship. JDI acknowledges support from an STFC Ernest Rutherford Fellowship (ST/W004119/1) and a University Academic Fellowship from the University of Leeds. Support for AFI was provided by NASA through the NASA Hubble Fellowship grant No. HST-HF2-51532.001-A awarded by the Space Telescope Science Institute, which is operated by the Association of Universities for Research in Astronomy, Inc., for NASA, under contract NAS5-26555. GL has received funding from the European Union's Horizon 2020 research and innovation program under the Marie Sklodowska-Curie grant agreement No. 823823 (DUSTBUSTERS). CL has received funding from the European Union's Horizon 2020 research and innovation program under the Marie Sklodowska-Curie grant agreement No. 823823 (DUSTBUSTERS) and by the UK Science and Technology research Council (STFC) via the consolidated grant ST/W000997/1. CP acknowledges Australian Research Council funding via FT170100040, DP18010423, DP220103767, and DP240103290. DP acknowledges Australian Research Council funding via DP18010423, DP220103767, and DP240103290. GR acknowledges funding from the Fondazione Cariplo, grant no. 2022-1217, and the European Research Council (ERC) under the European Union’s Horizon Europe Research \& Innovation Programme under grant agreement no. 101039651 (DiscEvol). H-WY acknowledges support from National Science and Technology Council (NSTC) in Taiwan through grant NSTC 113-2112-M-001-035- and from the Academia Sinica Career Development Award (AS-CDA-111-M03). GWF acknowledges support from the European Research Council (ERC) under the European Union Horizon 2020 research and innovation program (Grant agreement no. 815559 (MHDiscs)). GWF was granted access to the HPC resources of IDRIS under the allocation A0120402231 made by GENCI. TCY acknowledges support by Grant-in-Aid for JSPS Fellows JP23KJ1008. Support for BZ was provided by The Brinson Foundation. This work was partly supported by the Deutsche Forschungsgemein- schaft (DFG, German Research Foundation) - Ref no. 325594231 FOR 2634/2 TE 1024/2-1, and by the DFG Cluster of Excellence Origins (www.origins-cluster.de). This project has received funding from the European Research Council (ERC) via the ERC Synergy Grant ECOGAL (grant 855130). Views and opinions expressed by ERC-funded scientists are however those of the author(s) only and do not necessarily reflect those of the European Union or the European Research Council. Neither the European Union nor the granting authority can be held responsible for them.  

\begin{appendix}

\section{Peak Intensity Maps}
\label{app:galleries}

\begin{figure*}
    \centering
    \includegraphics[width=\textwidth]{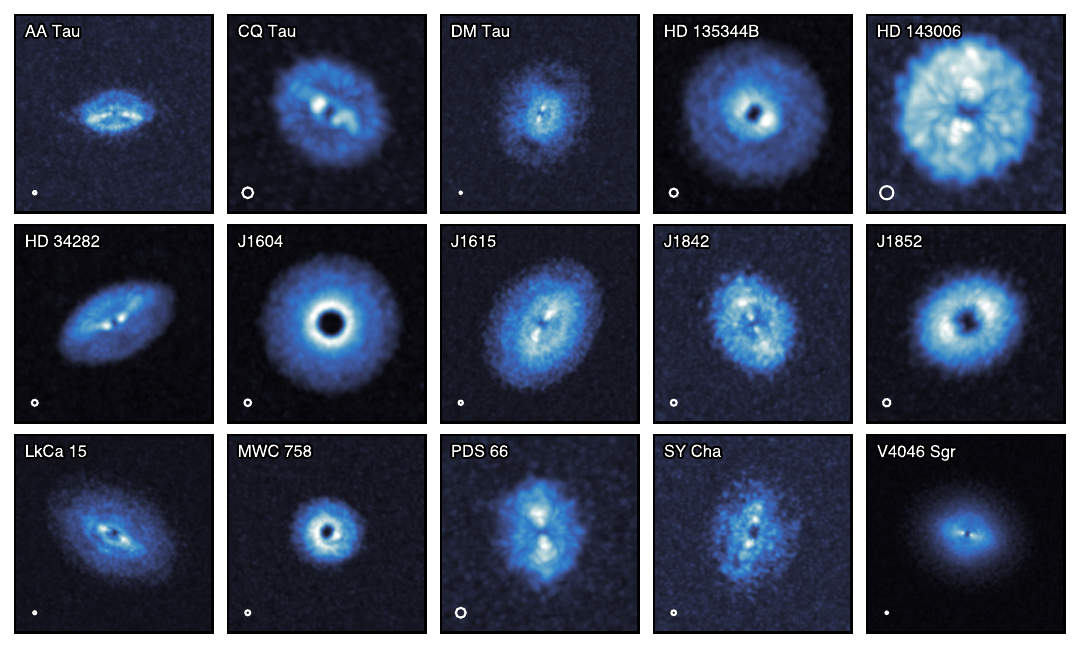}
    \caption{Gallery of $^{13}$CO J=3-2 emission from the exoALMA sources. The field of view of each panel is the same as that in Figure~\ref{fig:CO_gallery_main}. The $0\farcs15$ beam is shown in the lower left corner. The linear color scaling is set to span up to 95\% of the peak value in each panel.}
    \label{fig:13CO_gallery}
\end{figure*}

\begin{figure*}
    \centering
    \includegraphics[width=\textwidth]{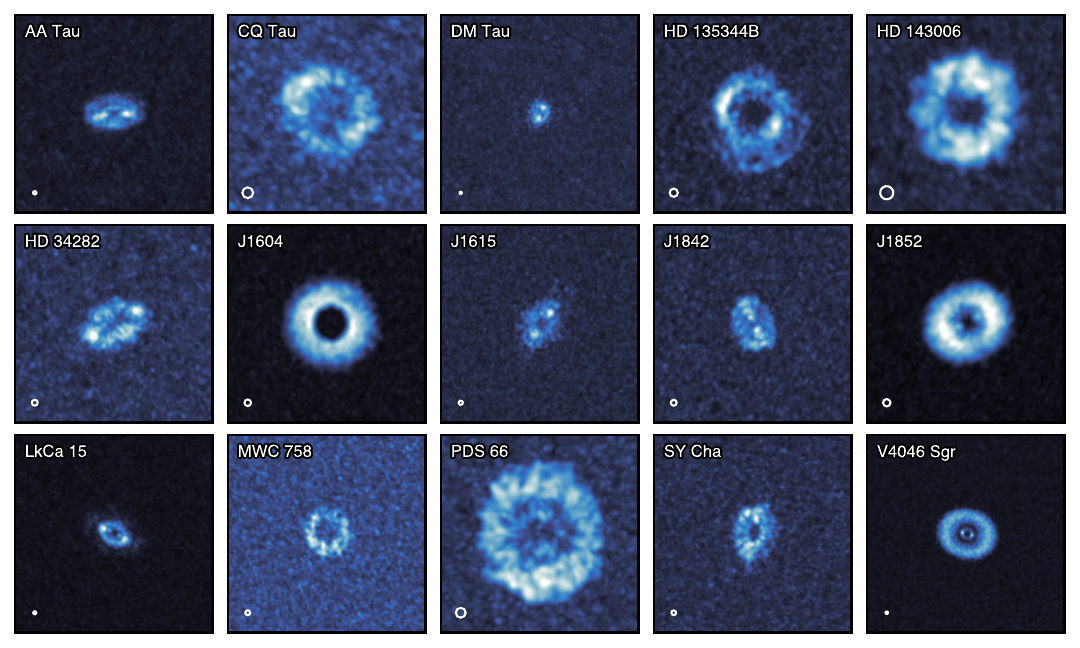}
    \caption{Peak intensity of the CS J=7-6 emission for all exoALMA sources. The field of view of each panel is the same as that in Figure~\ref{fig:CO_gallery_main}. The $0\farcs15$ beam is shown in the lower left corner. The linear color scaling is set to span up to 95\% of the peak value in each panel.}
    \label{fig:CS_gallery}
\end{figure*}

In this Appendix we present the peak intensity maps of the $^{13}$CO J=3-2 and CS J=7-6 emission for all the exoALMA sources in Figures~\ref{fig:13CO_gallery} and \ref{fig:CS_gallery}, respectively. As with Figure~\ref{fig:CO_gallery_main}, the `fiducial set' of images were used meaning a spatial resolution of $0\farcs15$ for all lines and a channel spacing of $100~{\rm ms\,s^{-1}}$ for $^{12}$CO and $^{13}$CO and a coarser $200~{\rm m\,s^{-1}}$ for the CS. The azimuthal asymmetries observed in the CS emission are due to radiative transfer effects arising from low optical depth emission that is in a ring morphology which results in a larger projected column density around the major axis of the disk. \texttt{bettermoments} was used for the peak extraction using a quadratic fit to the line profile. For a gallery of the continuum emission we refer the reader to \citet{Curone_exoALMA}.

\end{appendix}

\bibliography{main}{}
\bibliographystyle{aasjournal}

\end{document}